\documentclass[12pt]{article}
\usepackage{amsmath}
\usepackage[dvips]{graphicx}
\usepackage{epsfig}
\usepackage{amsmath}
\usepackage{amssymb}
\usepackage{graphics,epstopdf}
\usepackage{amsfonts}
\usepackage{amsthm}
\usepackage[usenames]{color}

\definecolor{AV}{rgb}{0.65,0.0,0}
\definecolor{GC}{rgb}{0,0.0,0.65}
\definecolor{WS}{rgb}{0,0.65,0}

\usepackage[
      colorlinks=true,
      linkcolor=blue,
      urlcolor=blue,
      filecolor=blue,
      citecolor=blue,
      pdfstartview=FitV,
      pdftitle={},
      pdfauthor={},
      pdfsubject={},
      pdfkeywords={},
      pdfpagemode=None,
      bookmarksopen=true
]{hyperref}

\newcommand{\bm}{\begin{multiline}}
\newcommand{\beq}{\begin{equation}}
\newcommand{\eeq}{\end{equation}}
\newcommand{\beqs}{\begin{eqnarray}}
\newcommand{\eeqs}{\end{eqnarray}}

\begin{document}

\thispagestyle{empty}

\hfill{}

\hfill{}

\hfill{}

\vspace{32pt}

\begin{center}

\textbf{\Large  A black ring on the Taub-bolt instanton in five dimensions}

\vspace{48pt}

\textbf{ Cristian Stelea,}\footnote{E-mail: \texttt{cristian.stelea@uaic.ro}}
\textbf{Marian C. Ghilea}\footnote{E-mail: \texttt{mghilea@gmail.com}}

\vspace*{0.2cm}

\textit{$^1$ Faculty of Physics, ``Alexandru Ioan Cuza" University}\\[0pt]
\textit{11 Bd. Carol I, Iasi, 700506, Romania}\\[.5em]

\textit{$^2$ Department of Physics, University of California}\\[0pt]
\textit{Santa Barbara, CA 93106-9530}\\[.5em]

\end{center}

\vspace{30pt}

\begin{abstract}
Using a solution generating technique, we derive a new exact solution describing a charged static black ring on the Taub-bolt gravitational instanton in five dimensions. Unlike the black ring constructed on the self-dual Taub-NUT instanton, it turns out that it is possible to find values of the parameters for which the static black ring is in equilibrium and the conical singularities disappear. We compute its conserved charges and discuss some of its thermodynamic properties.  
\end{abstract}

\vspace{32pt}

\setcounter{footnote}{0}

\newpage

\section{Introduction}

While the physics of four dimensional black holes is pretty much understood, in recent years one witnessed remarkable developments in the study of higher dimensional black holes (for a review see \cite{Emparan:2008eg,Obers:2008pj}). There are many motivations to embark in such a study. For instance, higher than four dimensions are required by string/M-theory, which is candidate for a unified field theory, aiming to provide us with a consistent theory of quantum gravity. Another motivation for studying gravity in higher dimensions is intrinsic, as a way to better understand the nature of the theory itself. In particular, one should be able to answer the question: which properties are characteristic to four dimensions and which are common to gravity in higher than four dimensions. One important advance in this direction was the Emparan and Reall's discovery of an exact solution in five dimensions describing an asymptotically flat rotating black ring \cite{Emparan:2001wn,Emparan:2001wk}. The black ring provided the first nontrivial example that known properties of the four dimensional black holes, such as the uniqueness theorem, do not hold in higher than four dimensions. Indeed, unlike their four dimensional counterparts, which necessarily posses a spherical horizon \cite{Hawking:1971vc} as a consequence of the topological censorship theorem \cite{Friedman:1993ty,Galloway:1999bp}, black holes in higher dimensions can have other horizon topologies as long they admit non-negative scalar curvature \cite{Galloway:1999br,Cai:2001su}. The black ring is a prime example of such an exotic black hole, as it has a ring topology $S^2\times S^1$ of the horizon, with rotation in the $S^1$ direction. Moreover, in certain conditions it can carry the same mass and angular momenta as the spherical Myers-Perry black hole \cite{Myers:1986un,Emparan:2004wy,Emparan:2006mm}. As a consequence, the uniqueness theorems for black holes in four dimensions cannot be extended to the five dimensional case without further assumptions of additional symmetry and specification of the rod structure \cite{Hollands:2007aj}. 

In the black ring solution the rotation along the $S^1$ direction creates a centrifugal force, which opposes the collapse of the black ring under its own gravity. If these forces are not in balance there are conical singularities in the space-time to keep the black ring in equilibrium. For a certain value of the angular momentum these competing forces will balance and the conical singularities disappear, rendering the solution regular on and outside the horizon. A more general solution describing a balanced rotation black ring, with rotation not only along $S^1$ but also along the azimuthal direction of $S^2$ has been presented in \cite{Pomeransky:2006bd}. This solution generalized the Emparan and Reall's black ring but also the black rings found in \cite{Mishima:2005id,Figueras:2005zp}, which had rotation only along the azimuthal direction of $S^2$. The general unbalanced black ring solution was subsequently found in \cite{Morisawa:2007di,Chen:2011jb} and, demanding the absence of the conical singularities, it reduces to the balanced rotating black ring. Following the discovery of the rotating black ring, its generalization to black Saturn \cite{Elvang:2007rd} and multi-black rings have been found in five dimensions \cite{Iguchi:2007is,Evslin:2007fv,Elvang:2007hs,Izumi:2007qx}. 

Black rings carrying electric/dipole charges have also been studied. For instance, a static black ring with an electric charge has been considered in \cite{Kunduri:2004da}. In that work is was found that the static black ring cannot be stabilized against its own gravitational collapse even in presence of an electric charge. However, the static charged black ring could be stabilized and the conical singularities eliminated if one immerses the system in a background electric field. The drawback of this construction was that the black ring is no longer asymptotically flat. Configurations involving charged black rings in various theories, including string theory, have been considered in \cite{Elvang:2003yy,Elvang:2003mj,Hoskisson:2008qq,Gal'tsov:2009da,Chng:2008sr,Stelea:2011jm}. Concentric supersymmetric black rings in five dimensions were first constructed in \cite{Gauntlett:2004wh,Gauntlett:2004qy}. Black rings in higher than five dimensions have also been found, albeit in numerical form \cite{Kleihaus:2012xh,Kleihaus:2010pr}.

Recently, Chen and Teo introduced in \cite{Chen:2010ih} a new class of five-dimensional black hole solutions, the so-called black holes on gravitational instantons. For these solutions, in absence of black holes, the background geometry is a direct product of a trivial time direction with a four dimensional Ricci flat gravitational instanton having $U(1)\times U(1)$ symmetry \cite{Chen:2010zu}. For the asymptotically flat black rings the spatial part of the background corresponds to four dimensional flat space. Another interesting case is provided by the so-called squashed black holes \cite{Dobiasch:1981vh,Gibbons:1985ac,Rasheed:1995zv,Larsen:1999pp}, for which the corresponding spatial part of the background is given by the self-dual Taub-NUT gravitational instanton. The asymptotic geometry is the same with that encountered in the case of the Kaluza-Klein magnetic monopole \cite{Sorkin:1983ns,Gross:1983hb}. Given the properties of the self dual Taub-NUT instanton, black holes in such backgrounds look five-dimensional in the near-horizon region, while asymptotically they resemble four-dimensional objects with a compactified fifth dimension. A static black ring in this background has been constructed in \cite{Ford:2007th} and subsequently generalized to a rotating black ring in \cite{Camps:2008hb}. The most general double-rotating black ring on the Taub-NUT instanton was recently found in \cite{Chen:2012zb}. If one considers black holes/rings carrying Maxwell charges, there is a plethora of solutions  (see for instance \cite{Ishihara:2005dp,Nakagawa:2008rm,Tomizawa:2008rh,Stelea:2008tt,Tomizawa:2008qr,Gal'tsov:2008sh,Mizoguchi:2011zj,Iguchi:2011qi,Ishihara:2006iv,Elvang:2005sa,Matsuno:2012hf,Stelea:2009ur,Stelea:2011fj}). The squashed black holes have also been generalized to include Yang-Mills fields in \cite{Brihaye:2006ws}. Similarly to the case of asymptotically flat black holes/rings, one can prove uniqueness theorems for KK black holes by assuming additional symmetry and specification of other invariants \cite{Hollands:2008fm}. There are also known solutions describing extremal black holes on the Eguchi-Hanson \cite{Ishihara:2006pb,Tomizawa:2008tj,Ishihara:2006ig,Yoo:2007mq,Matsuno:2007ts} and also the Atiyah-Hitchin instanton \cite{Ghezelbash:2009zz,Stotyn:2008fk}.

Now, instead of the self-dual Taub-NUT instanton one could use the so-called Taub-bolt solution \cite{Page}, which is a non-self-dual asymptotically locally flat gravitational instanton (with the same asymptotics as the Taub-NUT instanton) while having a space-like direction with finite norm at infinity. Both the Taub-NUT and the Taub-bolt solutions can be obtained from the Lorentzian Taub-NUT solution by a Wick rotation of the time coordinate and also an analytical continuation of the nut parameter. The solution describing a static black hole on Taub-bolt instanton was recently discovered by Chen and Teo \cite{Chen:2010ih}. Its thermodynamics, including also an electric charge, has been investigated in \cite{Nedkova:2011hx}. This single black hole solution has been recently generalized to a system of two non-extremal black holes sitting at the two turning points of the Taub-bolt geometry \cite{Stelea:2012ph}. Surprisingly, the conical singularities  in this double black hole solution can be completely eliminated by careful choices of the various parameters and the double black hole system remains in equilibrium.

In this paper we are interested in constructing a five dimensional static black ring solution on the Taub-bolt instanton. The solution generating method that we shall use has been previously applied to construct multi-black hole objects in spaces with Kaluza-Klein asymptotics \cite{Stelea:2009ur,Stelea:2011fj,Stelea:2012ph}. The main idea of this method is to map a general static electrically charged axisymmetric solution of Einstein-Maxwell theory in four dimensions to a five-dimensional static electrically charged axisymmetric solution of the Einstein-Maxwell system. The final five dimensional solution is determined up to the choice of a harmonic function. By carefully choosing the form of this function, one can construct the appropriate rod structures to describe the desired configuration of a black holes/rings on the Taub-bolt instanton. For example, as it was shown in \cite{Stelea:2012ph}, starting from the four dimensional Reissner-Nordstr\"om solution one is able to recover the solution describing a five dimensional black hole sitting at one of the turning points of the Taub-bolt instanton. Another simple choice of the harmonic functionl leads us to the desired solution describing a five dimensional static and charged black ring on the Taub-bolt instanton. The main result of this paper is showing that in the Taub-bolt background one can keep this static black ring in equilibrium, that is, it is possible to completely eliminate the conical singularities in this system. To our knowledge, this is the first example of a regular static black ring solution in spaces with Kaluza-Klein asymptotics.

The structure of this paper is as follows: in the next section we derive the solution describing a static charged black ring on the Taub-bolt background. In section $3$ we discuss the rod structure of this solution and compute its conserved charges. Finally, we end with a summary of our work and avenues for further research.

\section{The black ring on the Taub-bolt instanton}

 Start with the static electrically charged black hole solution written here in Weyl form:
\begin{eqnarray}  \label{RN4dim}
ds^2&=&-\tilde{f}dt^2+\tilde{f}^{-1} \big[e^{2\tilde{\mu}}(d\rho^2+dz^2)+%
\rho^2 d\varphi^2\big], \\
\Psi&=&-\frac{4q}{r_2+r_3+2m},~~~~~~~\tilde{f}=\frac{(r_{2}+r_{3})^2-4%
\sigma^2}{(r_{2}+r_{3}+2m)^2},~~~~~~~ e^{2\tilde{\mu}}=\frac{Y_{23}}{2r_{2}r_{3}},  \notag
\end{eqnarray}
where we define in general
\begin{eqnarray}
r_{i}&=&\sqrt{\rho^2+\zeta_i^2}, ~~~~~~~ \zeta_i=z-a_i, ~~~~~ Y_{ij}=r_ir_j+\zeta_i\zeta_j+\rho^2, 
\end{eqnarray}
while here $a_2=-\sigma$, and $a_3=\sigma$. Note that $\sigma=\sqrt{m^2-q^2}$, where $m$ denotes the mass and $q$ the electric charge of the black hole, while $\Psi$ is the electric potential in four dimensions. Then the corresponding solution of the Einstein-Maxwell system in five dimensions with Lagrangian
\begin{eqnarray}
\mathcal{L}_{5}=\sqrt{-g}\left[R -\frac{1}{4}F_{(2)}^2\right]
\label{EMDaction5d}
\end{eqnarray}
where $F_{(2)}=dA_{(1)}$ can be written as:
\beqs
\label{final5dalpha}
ds_{5}^{2}&=&-\tilde{f}dt^{2}+\tilde{f}^{-\frac{1}{2}}\bigg[\frac{e^{2h}}{A^2-C^2e^{4h}}(d\chi+4ACH d\varphi)^{2}+(A^2-C^2e^{4h})e^{\frac{3\tilde{\mu}}{4}
+2\gamma-2h}(d\rho ^{2}+dz^{2})\nonumber\\
&&+\rho^2(A^2-C^2e^{4h})e^{-2h}d\varphi ^{2}\bigg],~~~~~~~
A_{(1)}=\frac{\sqrt{3}}{2}\Psi dt.
\eeqs
Here $A$ and $C$ are constants, while $h$ is an harmonic function, which is so far arbitrary. By carefully choosing the form of $h$, one can construct the appropriate rod structure to describe the wanted configuration of a black ring on the Taub-bolt instanton. In this case, the appropriate choice turns out to be:
\beqs
e^{2h}&=&\frac{r_2+\zeta_2}{r_1+\zeta_1}\sqrt{\frac{r_3+\zeta_3}{r_2+\zeta_2}}\frac{r_4+\zeta_4}{r_3+\zeta_3}\equiv\sqrt{\frac{r_2+\zeta_2}{r_3+\zeta_3}}\frac{r_4+\zeta_4}{r_1+\zeta_1},
\eeqs
where $-a_1=R_1>\sigma$ and $a_4=R>\sigma$. The second factor corresponds to a `correction' that has to be taken into account for the black hole horizon in the four-dimensional seed solution, while the first and the third factors simply correspond to finite rods along the $\chi$-direction, the first one which starts at $a_1$ and ends at $a_2$, while second one starts at $a_3$ and ends at $a_4$.

Once the form of $h$ has been specified for this particular solution, the remaining function  $\gamma$ can be obtained by quadratures using the equations:
\begin{eqnarray}  \label{gammap1a}
\partial_\rho{\gamma}&=&\rho[(\partial_\rho h)^2-(\partial_z h)^2],~~~~~~~
\partial_z{\gamma}=2\rho(\partial_\rho h)(\partial_z h).
\end{eqnarray}
 By integrating (\ref{gammap1a}) one simply obtains:
\beqs
e^{2\gamma}&=&\frac{\sqrt{2\sqrt{2}}}{K_0}\frac{Y_{14}}{r_1r_4}\left(\frac{Y_{12}Y_{34}}{Y_{13}Y_{24}}\right)^{\frac{1}{2}}\left(\frac{Y_{23}}{r_2r_3}\right)^{\frac{1}{4}},
\eeqs
where $K_0$ is a constant to be fixed later.

Also, the function $H$ is the so-called `dual' of $h$\footnote{By this we mean that it is a solution of $dH=\rho(\partial_{\rho}h dz-\partial_zhd\rho)$. In general, the dual of $\frac{1}{2}\ln(r_i+\zeta_i)$ is $-\frac{1}{2}(r_i-\zeta_i)$, while the dual of $\frac{1}{2}\ln(r_i-\zeta_i)$ is $-\frac{1}{2}(r_i+\zeta_i)$, where $r_i=\sqrt{\rho^2+\zeta_i^2}$, $\zeta_i=z-a_i$ and $a_i$ is constant.} and it is easily evaluated to be:
\beqs
H&=&\frac{1}{4}\big[r_3-\zeta_3-r_2+\zeta_2+2(r_1-\zeta_1-r_4+\zeta_4)\big]=\frac{1}{4}\big[r_3-r_2+2(r_1-r_4)\big]
\eeqs
up to a additive constant factor. Then the final solution in five dimensions can be written in the form:
\beqs
ds^2&=&-\tilde{f}dt^2+\frac{F}{\Sigma}(d\chi+\omega d\varphi)^2+\frac{\Sigma}{G}\big[e^{2\mu}(d\rho^2+dz^2)+\rho^2 d\varphi^2\big],\nonumber\\
A_{(1)t}&=&\frac{\sqrt{3}}{2}\Psi,
\eeqs
where we defined the following functions:
\beqs
F&=&\tilde{f}^{-\frac{1}{2}}e^{2h},~~~~G=\tilde{f}^{\frac{1}{2}}e^{2h},~~~~\Sigma=A^2-C^2e^{4h},~~~~e^{2\mu}=e^{\frac{3\tilde{\mu}}{2}+2\gamma},~~~~\omega=4ACH.
\label{functions}
\eeqs
In the following section we shall prove that for certain values of the parameters $A$ and $C$ this solution describes a regular static charged black ring on the Taub-bolt instanton.

\section{Properties of the black ring solution}

So far the constants $A$ and $C$ were kept arbitrary. However, in order to have the right asymptotics at infinity, it turns out that one has to impose the following condition $A^2=C^2+1$. This follows from asking that $\frac{F}{\Sigma}\rightarrow 1$ in the asymptotic region.

Let us consider the rod structure of the above solution. Following the procedure outlined in \cite{Chen:2010zu,Harmark:2004rm}, note that there are four turning points that divide the $z$-axis into five rods as follows:\footnote{We are writing the vectors in the basis $\{\partial/\partial t,  \partial/\partial \chi, \partial/\partial\varphi\}$.} 
\begin{itemize}
\item For $z<a_1$ one has a semi-infinite space-like rod, with normalized direction: 
\beqs
l_1&=&\sqrt{\frac{4}{K_0}}(0,2AC(R+R_1-2\sigma),1).
\eeqs
\item For $a_1<z<a_2$ one has a finite space-like rod. Its normalized rod direction is given by $l_2=\frac{1}{k_2}(0,1,0)$, where 
\beqs
k_2&=&\frac{1}{2A^2}\sqrt{\frac{K_0}{4}}\frac{\sqrt{R_1+\sigma}}{(R+R_1)\sqrt{R_1-\sigma}}
\eeqs 
is the Euclidian surface gravity on this rod.
\item For $a_2<z<a_3$ one has a finite time-like rod, corresponding to the black ring horizon, having the normalized direction $l_3=\frac{1}{k_{E}}(1,0,0)$, where 
\beqs
k_{E}&=&\frac{1}{A}\sqrt{\frac{K_0}{4}}\frac{1}{(R+R_1)(m+\sigma)}\sqrt{\frac{\sigma(R_1+\sigma)(R+\sigma)}{2(m+\sigma)}}
\eeqs
is the surface gravity of the black ring.
\item For $a_3<z<a_4$ one has again a finite space-like rod. Its normalized rod direction is given by $l_4=\frac{1}{k_4}(0,1,0)$, where 
\beqs
k_4&=&\frac{1}{2A^2}\sqrt{\frac{K_0}{4}}\frac{\sqrt{R+\sigma}}{(R+R_1)\sqrt{R-\sigma}}
\eeqs 
is the Euclidian surface gravity on this rod.
\item For $z>a_4$ one has an semi-infinite space-like rod with normalized direction 
\beqs
l_5=\sqrt{\frac{4}{K_0}}(0,-2AC(R+R_1-2\sigma),1).
\eeqs
\end{itemize}

\begin{figure}[tbp]
\par
\begin{center}
\includegraphics{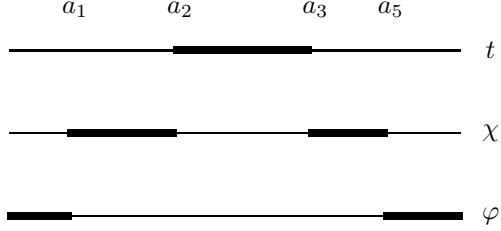} 
\end{center}
\caption{Rod structure of black ring on Taub-bolt.}
\label{fig}
\end{figure}

To have a proper black ring solution, it is necessary that the rod directions $l_2$ and $l_3$ be parallel as in Figure \ref{fig}. This can be easily achieved if one takes $R_1=R$. Finally, let us pick the value of the constant $K_0=4$. Then, if one defines the nut charge $n=2AC(R-\sigma)$, to obtain the rod structure that corresponds to the Taub-bolt instanton \cite{Chen:2010zu}, one has to impose the following condition:
\beqs
C&=&\frac{AR}{2\sqrt{R^2-\sigma^2}}.
\eeqs
Similarly to the case of a single black hole on the Taub-bolt instanton, if one parameterizes the constants $A^2=\frac{1}{1-\alpha^2}$ and $C^2=\frac{\alpha^2}{1-\alpha^2}$ such that $A^2=C^2+1$ is trivially satisfied for $\alpha<1$, then the nut parameter can be written as $n=\frac{2\alpha(R-\sigma)}{1-\alpha^2}$, while the parameter $\alpha$ has to satisfy the simple condition:
\beqs
\alpha&=&\frac{R}{2\sqrt{R^2-\sigma^2}}<1.
\label{alpha}
\eeqs

Then the final rod structure becomes:
\beqs
l_1&=&(0,2n,1),~~~l_2=(0,4n,0),~~~l_3=\frac{1}{k_E}(1,0,0),~~~l_4=(0,4n,0),~~~l_5=(0,-2n,1),\nonumber
\eeqs
where the black ring surface gravity is:
\beqs
k_E&=&\frac{1}{4R(m+\sigma)}\sqrt{\frac{\sigma(R+\sigma)(3R^2-4\sigma^2)}{2(R-\sigma)(m+\sigma)}}.
\eeqs
Note that there is the constraint $R>\frac{2\sigma}{3}$ to keep $\alpha<1$. Finally, to ensure regularity of the background geometry, the following identifications of the coordinates $(\chi,\varphi)$ have to be made \cite{Chen:2010ih}:
\beqs
(\chi,\varphi)\rightarrow (\chi+4n\pi,\varphi+2\pi),~~~(\chi,\varphi)\rightarrow(\chi+8n\pi,\varphi).
\eeqs
In conclusion, we have shown that the static black ring solution on the Taub-bolt instanton can become nonsingular, as long as one constrains the parameters as above. 

\subsection{Particular limits}

In the uncharged case, one simply takes $q=0$, that is $\sigma=m$. One should note that the uncharged black ring is still free of conical singularities as long as condition (\ref{alpha}) is satisfied. 

Another limit of interest is when the black ring is removed from the Taub-bolt instanton. This limit corresponds to taking $m=\sigma=0$. Then $A^2=\frac{4}{3}$, while $C^2=\frac{1}{3}$ and the nut charge becomes $n=\frac{4R}{3}$. It can be easily shown that the five-dimensional geometry can be cast in a Taub-NUT-like form with a trivial time direction:
\beqs
ds^2&=&-dt^2+f(r)(d\chi+2n\cos\theta d\varphi)^2+\frac{dr^2}{f(r)}+(r^2-n^2)(d\theta^2+\sin^2\theta d\varphi^2),\nonumber\\
f(r)&=&\frac{r^2-2Mr+n^2}{r^2-n^2}
\label{TN}
\eeqs
after performing the coordinate transformations:
\beqs 
\rho&=&\sqrt{r^2-2Mr+n^2}\sin\theta,~~~~~z=(r-M)\cos\theta,
\eeqs
as long as one takes $R=\sqrt{M^2-n^2}$. According to the previous discussion, the five-dimensional background is regular if the nut charge is $n=\frac{4R}{3}$. This constrains the value of the Taub-NUT mass to be $M=\frac{5|n|}{4}$, which is the regularity condition of the Taub-bolt instanton. In conclusion, the four-dimensional Taub-NUT-like part of the geometry corresponds indeed to that of the Taub-bolt instanton.

Finally, in the extremal limit, one should take $q=m$, that is $\sigma=0$ and the black ring horizon degenerates to a single point. The four-dimensional part of the metric can be cast into a Taub-NUT-like form with $R=\sqrt{M^2-n^2}$. The five dimensional extremal solution becomes:
\beqs
ds^2&=&-\frac{dt^2}{\left(1+\frac{m}{R_0}\right)^2}+\left(1+\frac{m}{R_0}\right)\big[f(r)(d\chi+2n\cos\theta d\varphi)^2+\frac{dr^2}{f(r)}+(r^2-n^2)(d\theta^2+\sin^2\theta d\varphi^2)\big],\nonumber\\
A_{(5)t}&=&-\frac{\sqrt{3}}{2}\frac{1}{1+\frac{m}{R_0}},
\eeqs
where we defined:
\beqs
R_0&=&\sqrt{(r-M)^2-(M^2-n^2)\sin^2\theta},
\eeqs
while $f(r)$ is the same as in (\ref{TN}). For  $M=\frac{5|n|}{4}$ the four dimensional geometry in the right bracket is precisely that of the Taub-bolt instanton. A preliminary analysis shows that the horizon of the extremal black ring degenerates as the $S^1$ direction of the black ring collapses to zero in this limit, such that there is a curvature singularity on the horizon. This resembles the situation of an extremal black ring in a asymptotically flat background.

\subsection{Conserved charges and thermodynamics}

The asymptotic region is found after performing the coordinate transformations:
\beqs
\rho=r\sin\theta,~~~~~~z=r\cos\theta.
\eeqs
and taking the limit $r\rightarrow\infty$. Since the asymptotic region is the same with that of the Kaluza-Klein magnetic monopole, to compute the conserved charges the easiest way is to use the counter-terms method, as described for instance in \cite{Mann:2005cx} or \cite{Kleihaus:2009ff}.

One obtains the conserved mass ${\cal M}$ and the gravitational tension ${\cal T}$ in the form:
\beqs
{\cal M}&=&\frac{L}{4G}\big[3m+(2R-\sigma)(1+2C^2)\big],~~~~~{\cal T}=\frac{(2R-\sigma)(1+2C^2)}{2G},
\eeqs
where $L=8\pi n$ is the length of the $\chi$ circle at infinity, while $G$ is the gravitational constant in five dimensions.

The electric charge is easily evaluated using Gauss formula, with the result ${\cal Q}=\frac{\sqrt{3}Lq}{4G}$, while the electric potential on the horizon is found to be:
\beqs
\Phi_H&=& -A_{(1)t}|_{horizon}=\frac{\sqrt{3}q}{m+\sigma}.
\eeqs

Note that ${\cal Q}\Phi_H=\frac{3(m-\sigma)L}{4G}$, upon using the relation $q^2=m^2-\sigma^2$. Finally, the area of the black ring horizon is computed to be:
\beqs
A_{H}&=&8\pi \sigma LA\frac{R(m+\sigma)}{R+\sigma}\sqrt{\frac{2(m+\sigma)}{\sigma}}.
\eeqs
and the entropy of the black ring is $S=\frac{A_H}{4G}$. The Hawking temperature is easily evaluated using the surface gravity of the horizon $k_E$. In this particular case one obtains:
\beqs
T_H&=&\frac{k_E}{2\pi}=\frac{1}{4\pi A}\frac{R+\sigma}{R(m+\sigma)}\sqrt{\frac{\sigma}{2(m+\sigma)}}.
\eeqs

It is now easy to verify that the following Smarr relation is satisfied:
\beqs
2{\cal M}-{\cal T}L&=&3T_HS+2{\cal Q}\Phi_H.
\eeqs

\section{Summary}

In this work we derived an exact solution describing a static and charged black ring on the Taub-bolt instanton. The solution generating technique that we used has been previously employed in \cite{Stelea:2009ur,Stelea:2011fj,Stelea:2012ph} to obtain exact solutions describing general configurations of charged black holes in five dimensional spaces whose asymptotic geometry resembles that of the KK magnetic monopole. Generically, for multi-black hole systems, the presence of conical singularities is unavoidable, as they are needed to keep the system static. This is not an impediment in discussing the thermodynamic properties of such spaces, since it turns out that the gravitational action is still well-defined \cite{Gibbons:1979nf,Costa:2000kf,Herdeiro:2010aq,Herdeiro:2009vd,Stelea:2011fj}. In section $2$ we derived the solution describing a static charged black ring on the Taub-bolt instanton. In section $3$ we showed that one can pick the parameters such that the rod structure of the background is the same as that of the Taub-bolt instanton. Further imposing appropriate identifications on the coordinates the conical singularities are avoided and the solution becomes completely regular. To our knowledge, this is the first regular solution describing a static black ring in spaces with KK asymptotics, without using Kaluza-Klein bubbles to stabilize the ring. We studied particular limits of the obtained black ring solution. Finally, by using a counter-terms method we computed the conserved charges and showed that a Smarr relation is satisfied in this case.

As avenues for further work, it would be interesting to consider more general configurations, containing black holes and black rings. For example, one should be able to construct a solution describing a static black Saturn on the Taub-bolt instanton and study its properties. More generally, one should be able to derive a rotating black ring on the Taub-bolt instanton and also find rotating generalizations of the black Saturn and bi/di-rings systems. Another interesting solution would correspond to a black ring in the same Taub-bolt background, for which the $S^1$ direction corresponds to the compactified dimensions $\chi$. However, to find such a solution one should recourse to a seed solution that is not static. Work on these subjects is currently in progress and it will be reported elsewhere.

\vspace{10pt}

{\Large Acknowledgements}

The work of C. S. was financially supported by POSDRU through the POSDRU/89/1.5/S/49944 project. Marian C. Ghilea is supported by the U.S. Department of Energy under Grant No. DE-SC0004036.

\end{document}